# Structural defects in MBE-grown CdTe-based heterojunctions for photovoltaic applications

Karolina Wichrowska, Tadeusz Wosinski*, Jaroslaw Z. Domagala, Slawomir Kret, Sergij Chusnutdinow, and Grzegorz Karczewski

*Institute of Physics, Polish Academy of Sciences, Aleja Lotnikow 32/46, PL-02668 Warsaw, Poland*

**ABSTRACT:**

Structural defects in the *p*-ZnTe/*i*-CdTe/*n*-CdTe single-crystalline heterojunctions designed for photovoltaic applications have been investigated by transmission electron microscopy (TEM) and deep-level transient spectroscopy (DLTS). Lattice parameters and misfit strain in the undoped CdTe absorber layers of the heterojunctions, grown by the molecular-beam epitaxy technique on two different substrates, GaAs and CdTe, have been determined with high-resolution X-ray diffractometry. A dense network of misfit dislocations at the lattice-mismatched CdTe/GaAs and ZnTe/CdTe interfaces and numerous threading dislocations and stacking faults have been shown by the cross-sectional TEM imaging of the heterojunctions. The DLTS measurements revealed five deep-level traps in the heterojunctions grown on the GaAs substrates and only three of them in the heterojunctions grown on CdTe. One of the traps, showing the exponential capture kinetics of charge carriers, has been identified as associated with the double acceptor level of Cd vacancies in the CdTe absorber layers. All the other traps have been attributed to the electronic states of extended defects, presumably dislocations, on the grounds of their logarithmic capture kinetics. Two of these traps, displaying the largest values of their capture cross-section and the properties characteristic of bandlike electronic states, have been ascribed to the core states of dislocations. It is argued that they are most likely responsible for decreased lifetime of photo-excited carriers resulting in a low energy conversion efficiency of solar cells based on similarly grown heterojunctions.

* Corresponding author: *E-mail address*: wosin@ifpan.edu.pl (T. Wosinski)

## 1. Introduction

Enhancement of the energy conversion efficiency of photovoltaic solar cells belongs to the one of essential requirements for further increase in the electricity-producing industrial solar photovoltaics. In spite of a tremendous improvement in the efficiency of photovoltaic systems during the last decades [1], especially those based on heterostructures of semiconductor compounds, it still remains far below the theoretical limits [2]. Structural defects, present in semiconductor heterostructures despite significant advancements gained in the epitaxial-growth technologies, belong to the most important factors limiting the performance of photovoltaic devices. Those defects, like misfit and threading dislocations, stacking faults, and twin and antiphase boundaries, result from a strain in the epitaxial layers caused by different lattice parameters and thermal properties of the layers composing the heterostructures and originate from the interfaces [3]. In particular, the dislocations, with broken (dangling) bonds in their cores, can give rise to electronic levels in the semiconductor band-gap, which act as recombination centers or traps for photo-excited charge carriers in the absorber layer of solar cells and result in a degradation of the conversion efficiency of photovoltaic devices.

One of the most promising absorber material for thin-film solar cells is cadmium telluride (CdTe). It is the group II-VI compound semiconductor with the zinc-blende crystalline structure, a direct band gap of 1.514 eV at room temperature that is nearly optimally matched to the solar spectrum for photovoltaic energy conversion, and a high absorption coefficient, of about $10^6$ $cm^{-1}$, at the above band-gap photon energy.

In this study we have applied high-resolution X-ray diffractometry (HXRD), transmission electron microscopy (TEM) and deep-level transient spectroscopy (DLTS) for extensive investigations of ZnTe/CdTe *p*-*i*-*n* heterojunctions, designed for photovoltaic applications, epitaxially grown on GaAs substrates. For comparison, we have also investigated similar heterojunctions grown on much more expensive CdTe substrates. Structural quality, lattice parameters and misfit strain in the CdTe absorber layers were characterized with HXRD. The cross-sectional TEM imaging was utilized to analyze the microscopic perfection of the CdTe/GaAs and ZnTe/CdTe interfaces and structural defects present in the investigated heterojunctions. Electronic properties of deep-level defects, i.e. their thermal activation energies and capture cross-sections, were revealed with the DLTS technique. Additional measurements of the kinetics for capture of charge carriers into the defect states enabled to distinguish isolated point defects from extended defects. Partial results have already been published in recent conference papers [4,5]. Here we combine the earlier findings with the new ones and more thorough discussion of the results.

## 2. Experimental details

The investigated *p*-ZnTe/*i*-CdTe/*n*-CdTe heterojunctions were grown by the molecular-beam epitaxy (MBE) technique on two different substrates: (i) lattice mismatched by 14.6% (001)-oriented GaAs and (ii) lattice matched (001)-oriented CdTe. Firstly, a highly iodine doped, by using $ZnI_2$ source, *n*-type CdTe contact layer, of above 10 μm thickness and about $8\times10^{18}$ $cm^{-3}$ electron concentration, was grown. The contact layer was covered by a 2 μm thick undoped CdTe absorber layer and, subsequently, by a 1 μm thick nitrogen-plasma doped *p*-type ZnTe layer. The *p*-type ZnTe layer, doped to about $2\times10^{18}$ $cm^{-3}$, facilitates for preparing low-resistivity contacts to the *p*-type side of the junction and increases the utilized spectral range of the solar spectrum. Both the CdTe and ZnTe layers were grown under stoichiometric conditions. In the case of GaAs substrate, prior to the deposition of CdTe layers, the substrate was covered with thin, a few-monolayer-thick, undoped ZnTe layer (nucleation layer) to reduce the strong lattice mismatch and to stabilize the growth along the [001] crystallographic direction. For more details on the growth peculiarities of the heterojunctions grown on GaAs substarate and their photovoltaic properties see Ref. [6].

The HXRD measurements were performed at room temperature using a high-resolution Philips X'Pert MRD diffractometer equipped with X-ray mirror, four bounce Ge(220) asymmetric monochromator and Ge(220) three bounce analyzer in triple axis configuration. The structural quality, lattice parameters and misfit strain were evaluated from the measured $2\theta/\omega$ scans and reciprocal lattice maps for the symmetrical 004 and asymmetrical $\bar{3}\bar{3}5$ Bragg reflections of Cu K$\alpha_1$ radiation. X-ray reciprocal lattice maps were recorded by performing a series of $2\theta/\omega$ scans. The high-resolution TEM imaging of thin, transparent for electrons, cross sections prepared from the heterojunctions were performed in a Titan Cubed 80-300 transmission electron microscope operating at 300 kV.

The DLTS measurements, employing a SEMITRAP DLS-82E system operating at a frequency of 1 MHz with a rate window realized by lock-in detection [7,8], were carried out using mesa-type structures of the dimensions of about 2×2 mm$^2$ prepared from the heterojunctions by etching in bromine-methanol solution to remove the upper layers of *p*-type ZnTe and undoped CdTe. Ohmic contacts were prepared by electroless gold deposition from $AuCl_3$ solution to *p*-ZnTe on the top of the mesa structure and by indium soldering to the *n*-CdTe contact layer.

## 3. Experimental results and discussion

### *3.1. HXRD results*

CdTe layers epitaxially grown on GaAs substrates are subjected to biaxial compressive strain, which results from the distinctly larger lattice parameter of CdTe than that of the substrate. At the growth temperature this strain is predominantly released by the formation of misfit dislocations at the interface. However, during the temperature lowering after the growth the CdTe layer suffers from additional compressive strain caused by the difference between the thermal expansion coefficients of the layer ($4.8\times10^{-6}$ K$^{-1}$) and the GaAs substrate ($5.7\times10^{-6}$ K$^{-1}$) [9]. At room temperature this residual strain, which is of the order of $10^{-4}$ [10], results in the lowering of the layer cubic symmetry to the tetragonal one with different in-plane and out-of-plane lattice parameters. Neither of the two causes of strain is expected to appear for homoepitaxially grown CdTe layers.

Fig. 1 presents the HXRD diffraction peaks ($2\theta/\omega$ scans) for the symmetrical 004 (Fig. 1a) and asymmetrical $\bar{3}\bar{3}5$ (Fig. 1b) Bragg reflections measured for the CdTe layers grown on GaAs and CdTe substrates and for the CdTe substrate used in the latter case. Angular positions of the diffraction peaks for symmetrical and asymmetrical reflections were used to determine the out-of-plane, $a_{\perp}$, and in-plane, $a_{||}$, lattice parameters, respectively. The obtained $a_{\perp}$ and $a_{||}$ values are listed in Table 1. Surprisingly, for both the measured layers the $a_{\perp}$ parameter was larger than that of the CdTe substrate, in contradiction to the expectation of the same lattice parameters of the layer and substrate in the case of homoepitaxial growth. In the latter case the $a_{||}$ lattice parameters of the layer and substrate are roughly the same, within the experimental errors (also given in Table 1), implying that the layer was grown pseudomorphically strained to the substrate, under small compressive strain, with the out-of-plane lattice parameter of the layer larger than that of the CdTe substrate.

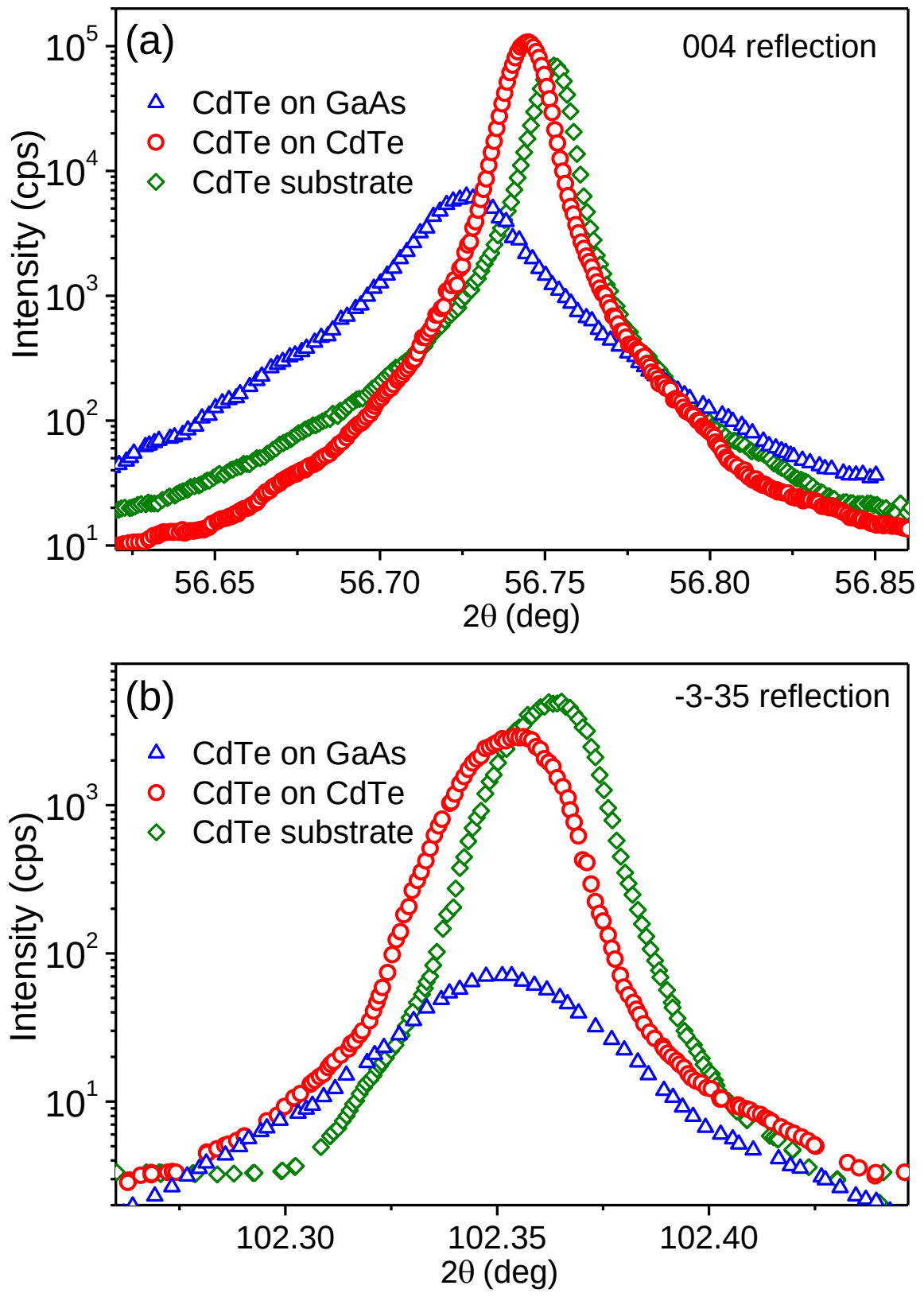

**Fig. 1.** HXRD $2\theta/\omega$ scans for the CdTe layers grown on GaAs and CdTe substrates and for the CdTe substrate obtained for the symmetrical 004 (a) and the asymmetrical $\bar{3}\bar{3}5$ (b) Bragg reflections. The presented scans have been smoothed for better clarity.

**Table 1.** The out-of-plane and in-plane lattice parameters for the CdTe layers grown on GaAs and CdTe substrates and for the CdTe substrate. The calculated relaxed lattice parameters and vertical strains for the layers grown on the two substrates are also shown.

| Layer (substrate) | $a_\perp$ (Å) (±0.0001) | $a_\parallel$ (Å) (±0.0003) | $a_{rel}$ (Å) | $\varepsilon_\perp$ ($\times 10^{-4}$) |
|---|---|---|---|---|
| CdTe (GaAs) | 6.4851 | 6.4811 | 6.4828 | 3.57 |
| CdTe (CdTe) | 6.4834 | 6.4817 | 6.4824 | 1.52 |
| CdTe substrate | 6.4822 | 6.4823 | – | – |

The relaxed lattice parameters for the epitaxial CdTe layers were calculated according to Eq. (1):

$$a_{rel} = \left(a_\perp + 2\frac{C_{12}}{C_{11}}a_\parallel\right) \Big/ \left(1 + 2\frac{C_{12}}{C_{11}}\right), \qquad (1)$$

where $C_{11} = 5.351\times10^{10}$ N/m$^2$ and $C_{12} = 3.681\times10^{10}$ N/m$^2$ are the elastic stiffness constants of CdTe [11]. The calculated $a_{rel}$ parameters are listed in Table 1. In addition, the vertical strain values in the layers, defined as: $\varepsilon_{\perp} = (a_{\perp} - a_{rel}) / a_{rel}$, were calculated and are also listed in Table 1.

The results presented in Fig. 1 and Table 1 point out that the CdTe layers, grown under compressive strain on GaAs substrate, are partially relaxed and display only residual vertical strain $\varepsilon_{\perp}$ of the order of $10^{-4}$ caused by different thermal expansion coefficients of CdTe and GaAs. Unexpectedly, the homoepitaxially grown CdTe layers display also a compressive vertical strain of the same order of magnitude. Interestingly, Heinke et al. [12] reported similar results for their CdTe layers homoepitaxially grown by the MBE technique on (001)-oriented CdTe substrates, which displayed even larger compressive vertical strain of about $5\times10^{-4}$.

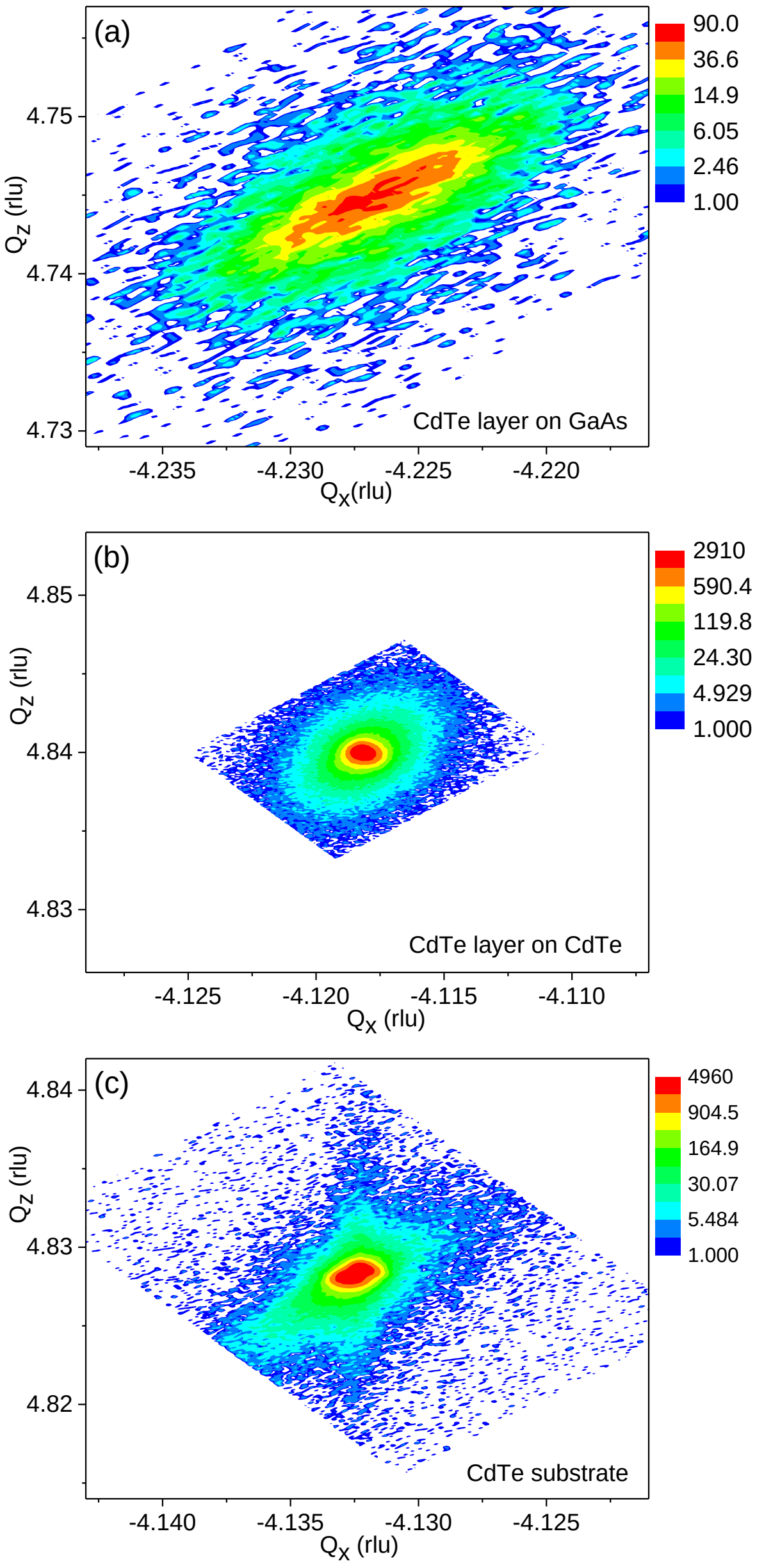

**Fig. 2.** Reciprocal lattice maps of the CdTe layers grown on GaAs (a) and CdTe (b) substrates and of the CdTe substrate (c) for the $\bar{3}\bar{3}5$ HXRD reflection. The horizontal and vertical axes correspond to the $2\theta/\omega$ and $\omega$ angles, respectively, in the reciprocal lattice units.

Reciprocal lattice maps of the CdTe layers grown on GaAs and CdTe substrates and of the CdTe substrate obtained for the asymmetrical $\bar{3}\bar{3}5$ Bragg reflection are presented in

Fig. 2. The results clearly demonstrate that the structural quality of the CdTe layer grown on CdTe substrate was significantly better than that of the layer grown on GaAs substrate and noticeably better that of the CdTe substrate. Moreover, the values of full width at half maximum (FWHM) of both the 004 and $\bar{3}\bar{3}5$ reflections of the measured intensity integrated along the $\omega$ vertical angle axis obtained for the homoepitaxially grown CdTe layer were lower than those for the CdTe substrate [4]. Interestingly, the relaxed lattice parameter of this layer is almost the same as the one obtained experimentaly for thick epitaxial CdTe layers by Fewster and Andrew [13] ($a_0$ = 6.482252(5) Å) from their very precise measurements with the use of high-resolution multiple-crystal multiple-reflection X-ray diffractometer, as shown in Table 1.

Heinke et al. [12] discussed possible reasons of the unusual compressive strain present in their homoepitaxially grown CdTe layers and suggested the twin formation as the most likely mechanism responsible for it. However, our thorough TEM investigations, described in the next Section, did not reveal the presence of any twins in the homoepitaxially grown CdTe layers. In addition, we have considered a possible influence of a high concentration of free charge carriers in highly *n*-doped CdTe contact layer on its lattice parameters, c.f. Ref. [14], but our calculated lattice changes due to the presence of free electrons turned out to be match smaller than the experimentally observed ones. Taking into account all these findings, we come to the conclusion that imperfect structural quality of the commercially available CdTe substrates could be responsible for the unexpected strain observed in homoepitaxially grown CdTe layers. Bridgman-grown bulk CdTe crystals usually contain a high concentration of cadmium vacancies [15], which may be the main reason of slightly decreased lattice parameters in the substrates used for homoepitaxial growth.

### *3.2. TEM results*

Cross-sectional TEM images of the CdTe/GaAs and ZnTe/CdTe interfaces of the heterojunctions grown on GaAs substrate are shown in Figs. 3 and 4, respectively. The strong lattice mismatch between CdTe and the GaAs substrate is accommodated by the formation of high density of misfit dislocations at the interface, as visible in Fig. 3. Generally, misfit dislocations act as a source of threading dislocations, which propagate through the epitaxial layer, as shown e.g. in Refs. [3] and [16]. Dense array of threading dislocations in the CdTe layer near the interface decreases their density with increasing distance from the interface (Fig. 3), but they are still present in the CdTe absorber layer up to the ZnTe/CdTe interface, as shown in Fig. 4. The lattice mismatch between ZnTe and CdTe of 6.2% results in tensile strain in the ZnTe layer, which is also accommodated by the formation of misfit dislocations at the interface and threading dislocations in the ZnTe layer (Fig. 4).

Complementary insight into the crystalline quality of the CdTe layers has been gained from selected area diffraction (SAD) patterns taken for different regions of the CdTe layers. Fig. 5 presents TEM-SAD patterns for the regions near the CdTe/GaAs interface with a dense array of dislocations (Fig. 5a and b) and near the ZnTe/CdTe interface with a rather low dislocation density (Fig. 5c). Two sets of Bragg reflections, from CdTe and ZnTe nucleation layer, are visible in Fig. 5a. In the zoomed part of this figure (Fig. 5b) higher order reflections are also visible. In addition, streaks-like diffuse scattering in between the Bragg reflections can be resolved in Fig. 5b. In contrast, the SAD pattern acquired from the upper part of the CdTe layer, near the ZnTe/CdTe interface, shown in Fig. 5c, presents the Bragg reflections from CdTe only. Moreover, no visible diffusion scattering features in Fig. 5c evidences for much higher crystalline quality of that part of the CdTe layer.

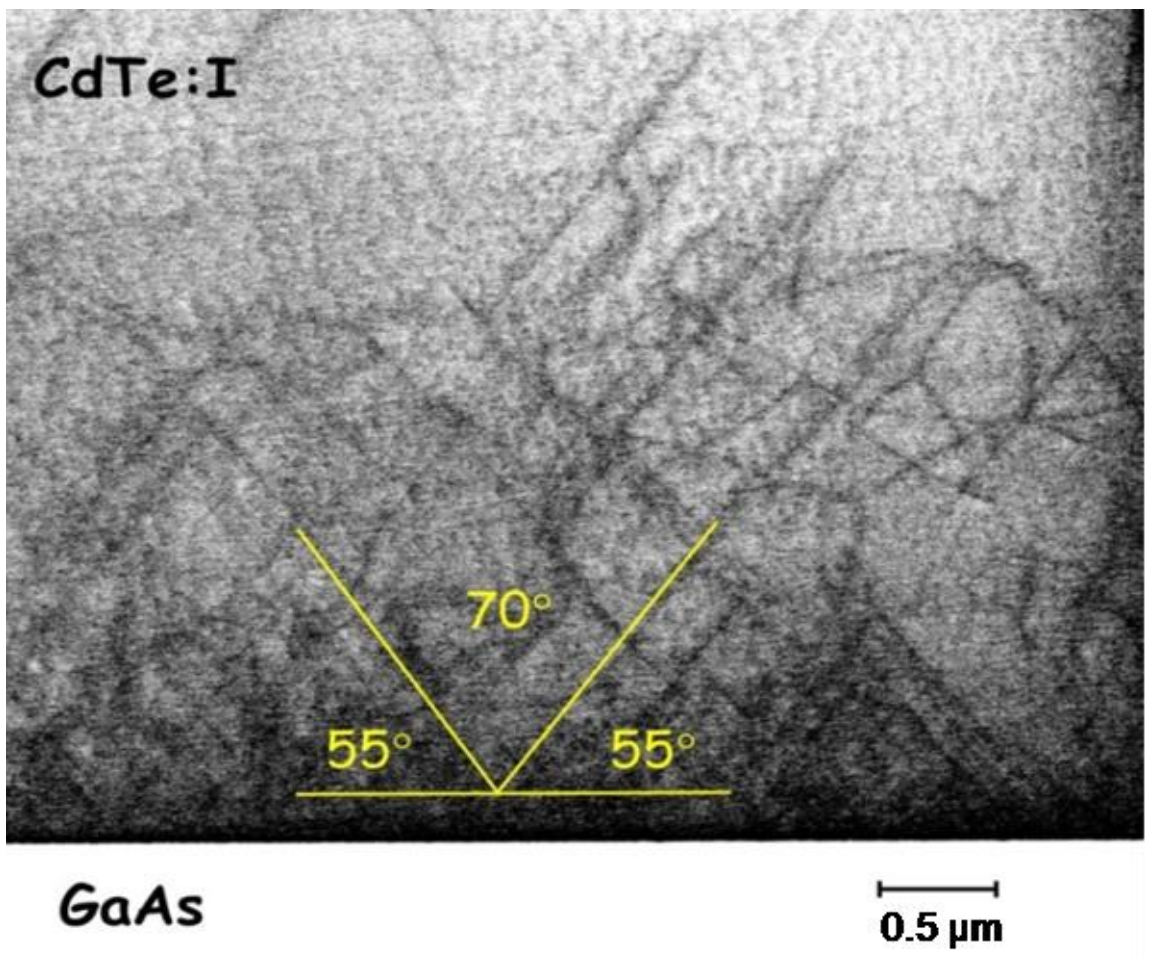

**Fig. 3.** Bright-field TEM image of the *n*-CdTe/GaAs interface in cross-section along the [110] zone axis for the *p*-ZnTe/*i*-CdTe/*n*-CdTe heterojuction grown on the GaAs substrate. Thin ZnTe nucleation layer between the GaAs substrate and CdTe layer is not visible.

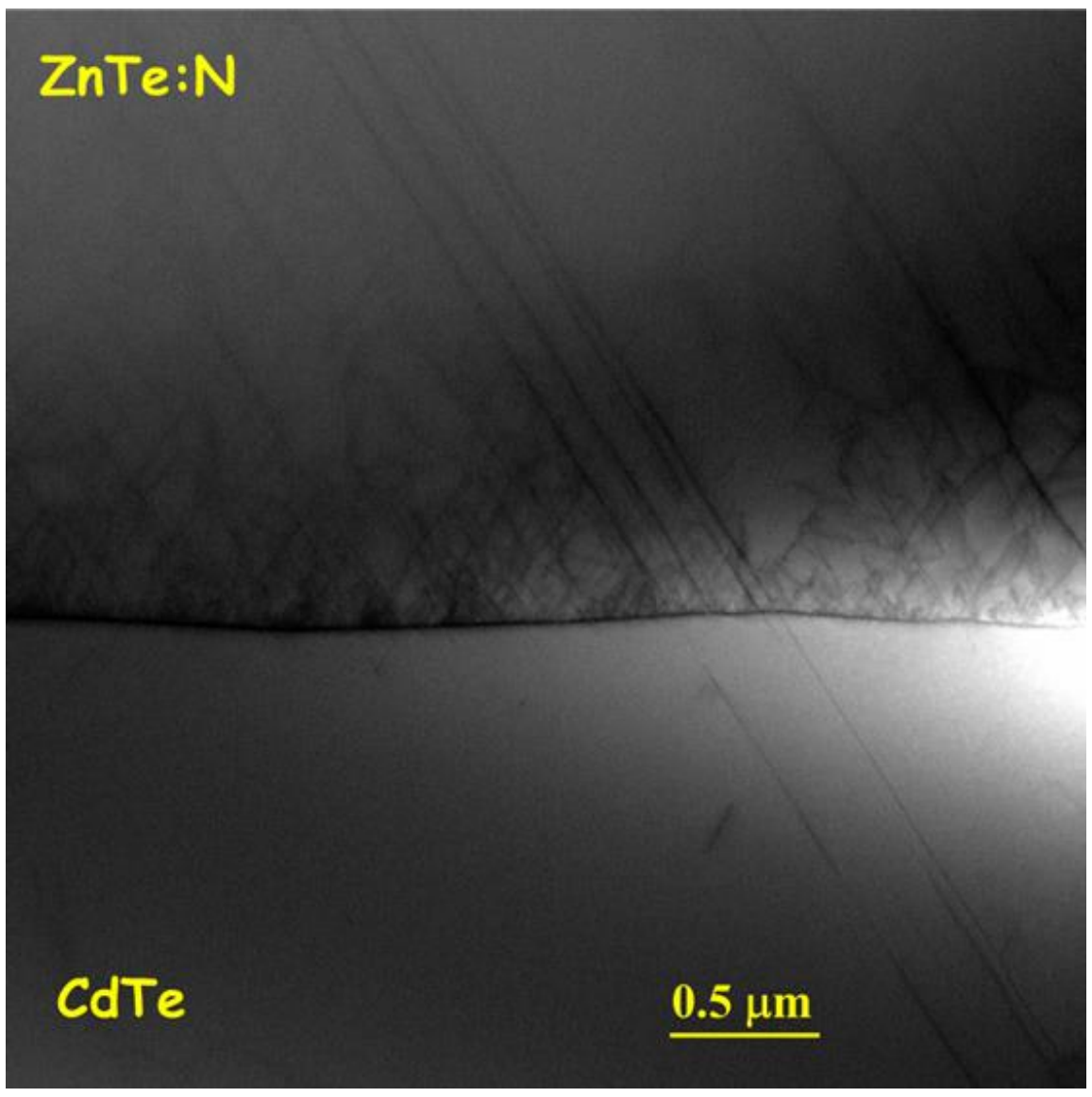

**Fig. 4.** Same as in Figure 3 for the *p*-ZnTe/*i*-CdTe interface.

Most of the threading dislocations, which segments intersecting the TEM foil are visible in Figs. 3 and 4, belong to two sets of dislocations. As depicted in Fig. 3, they are tilted at the angle of about 55° to the interface plane and display the angle of about 70° between the two sets of dislocations. The geometry of those dislocations points out that they are 60°-type dislocations aligned along various ⟨011⟩ crystallographic directions at the {111} glide planes, which display the angles of 54.74° with the (001) interface plane. 60°-dislocations are the most common dislocations in the zinc-blende crystalline structure and they may easily glide under residual stress through the epitaxial layers.

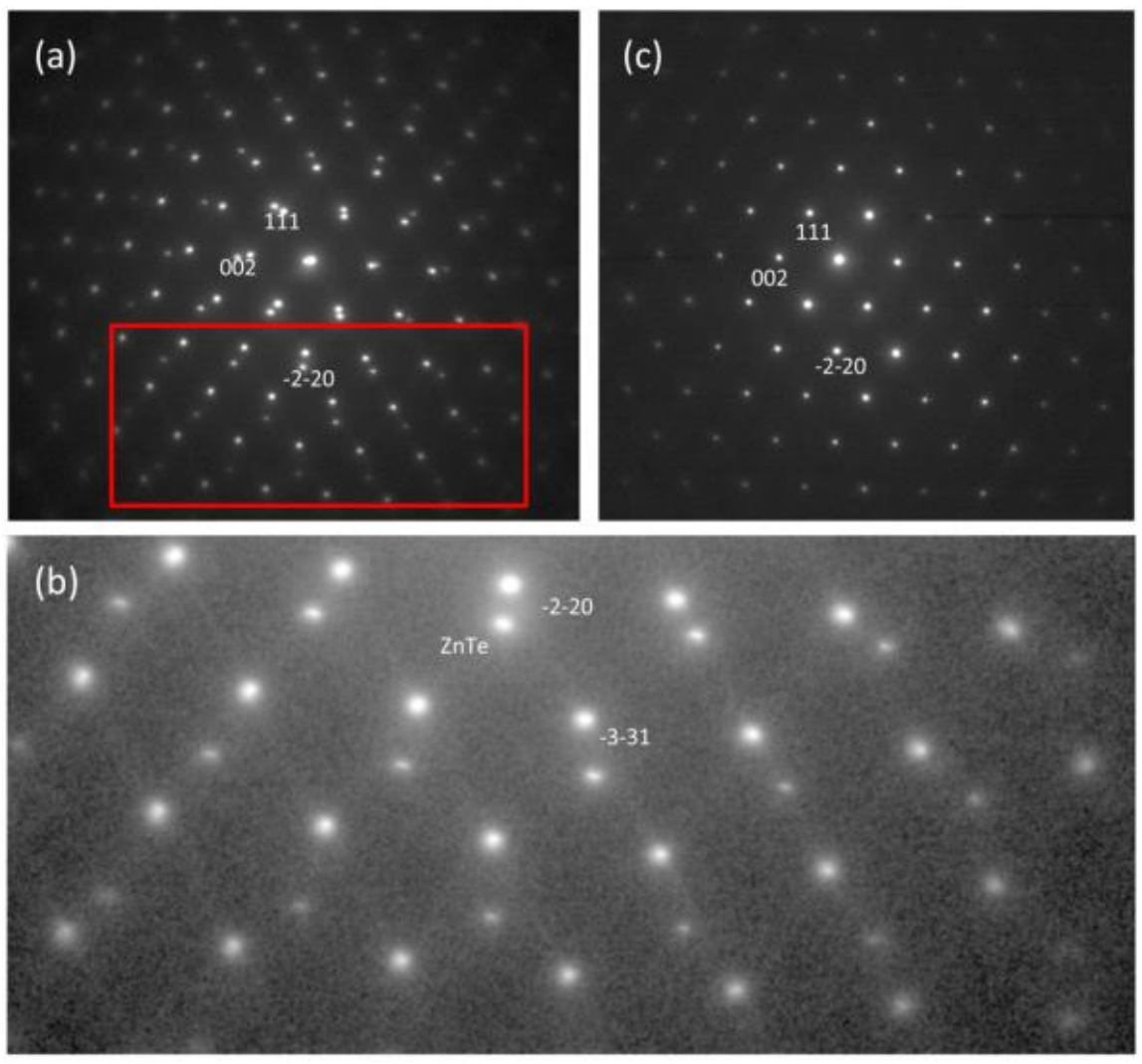

**Fig. 5.** TEM-SAD patterns acquired from two different areas, of about 600 nm in diameter, of the CdTe layer taken along the [110] zone axis. (a) SAD pattern acquired from the CdTe layer near the *n*-CdTe/GaAs interface shown in Figure 3. (b) zoomed part of the SAD pattern outlined with a red frame in (a). (c) SAD pattern acquired from the upper part of CdTe layer near the *p*-ZnTe/*i*-CdTe interface shown in Figure 4.

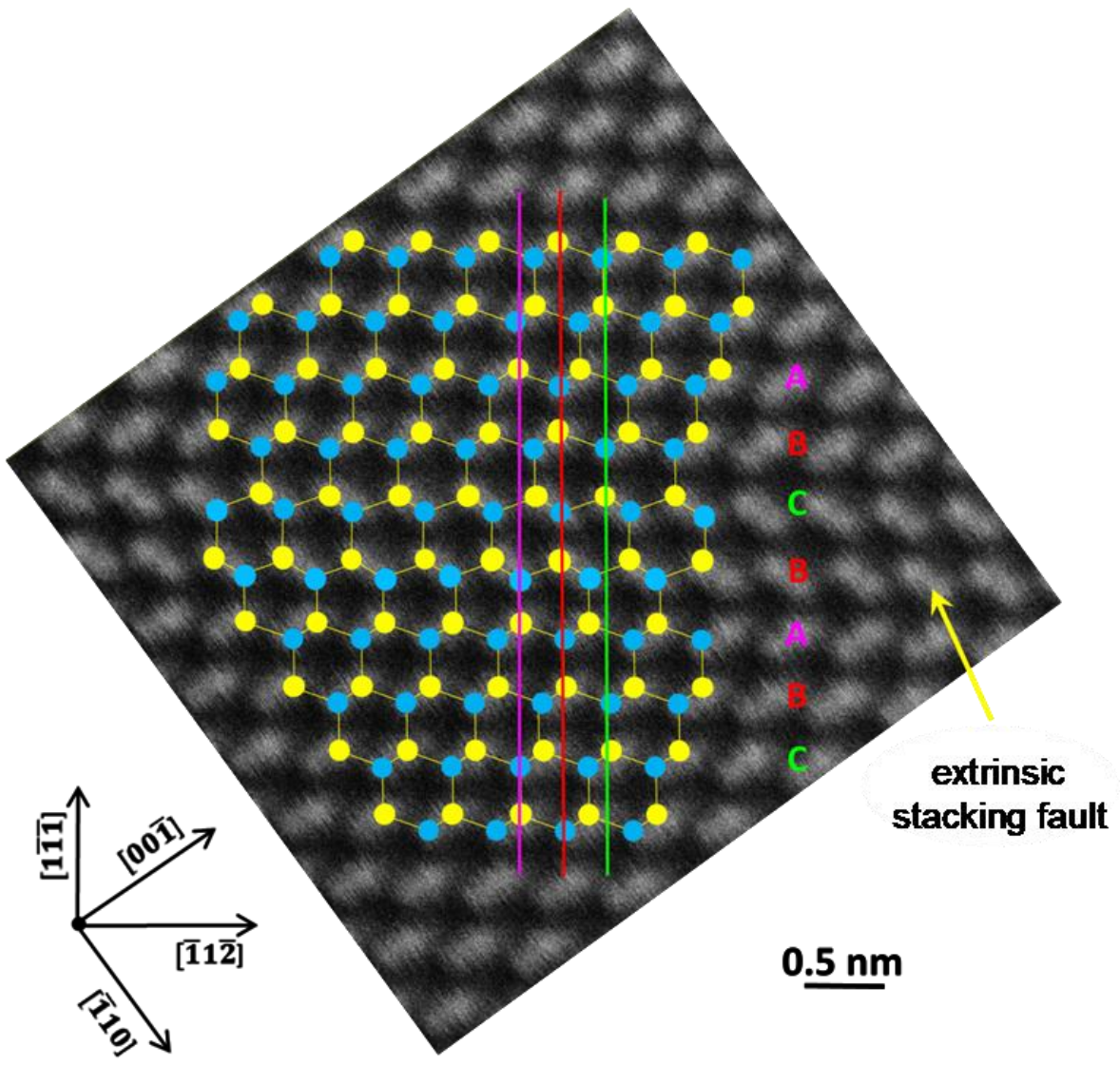

**Fig. 6.** Z-contrast TEM image showing extrinsic stacking fault in the *i*-CdTe absorber layer of the *p*-ZnTe/*i*-CdTe/*n*-CdTe heterojunction grown on the GaAs substrate.

The mismatch strain in the epitaxial layers leads also to the formation of a number of stacking faults, which are created by the dissociation of perfect dislocations into partial dislocations enclosed a stacking fault. An extrinsic stacking fault (two layers stacking fault) is created when one additional {111}-type close-packed plane is inserted into the perfect zinc-blende crystal. An example of such a stacking fault, revealed in a high-resolution TEM image of the CdTe layer grown on GaAs substrate, is shown in Fig. 6. On the other hand, an intrinsic stacking fault (one layer stacking fault) is created when one close-packed plane is removed from the crystal. An example of this type stacking fault revealed in the ZnTe layer is presented in Fig. 7. The two types of stacking faults were observed in both the CdTe and ZnTe layers in the heterojunctions grown on GaAs as well as in the ZnTe layers grown on the homoepitaxially grown CdTe layer.

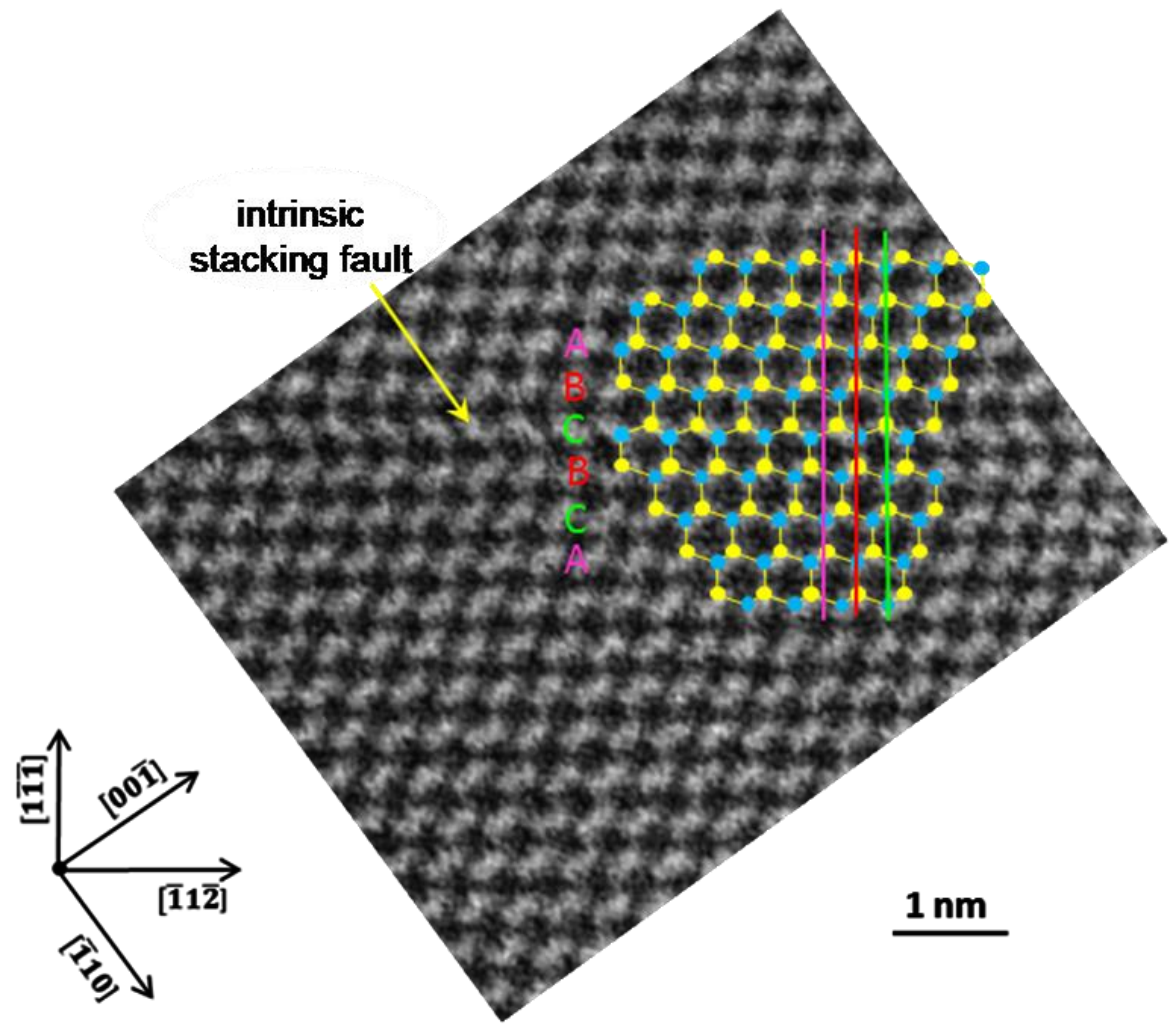

**Fig. 7.** Same as in Figure 6 showing intrinsic stacking fault in the *p*-ZnTe layer.

Both the types of stacking faults form very thin layers of wurtzite structure buried in the zinc-blende type host crystal. First-principles calculations predict that the presence of buried wurtzite layers, which behave like hole-trap centers in *p*-type zinc-blende CdTe, significantly affect the electronic properties of CdTe layers [17]. Moreover, Poplawsky et al. [18] have recently shown, by the use of electron-beam-induced current (EBIC) measurements combined with TEM structural analysis, that small grains of hexagonal wurtzite structure present in CdTe-based solar cells are not photoactive.

Importantly, thorough examination of our TEM images did not reveal any extended defects in the homoepitaxially grown CdTe layers. Structural perfection of these layers was substantially better than that of the layers grown on the GaAs substrate. On the other hand, the ZnTe layers grown on the homoepitaxially grown CdTe layers contained numerous dislocations and stacking faults, similarly as those shown in Figs. 4 and 7.

### *3.3. DLTS results*

Deep-level transient spectroscopy, which belongs to the junction spectroscopy techniques, is currently the most widely used one for characterizing electrically active defects in semiconductor materials and structures [19]. The DLTS technique, developed by Lang in 1974 [20,21], consists in the measurements of the capacitance changes of the space charge region in a reverse-biased Schottky or *p*-*n* junction as a result of charge-carrier trapping at deep-level defect states during the so-called filling voltage pulse. The technique provides so-called defect “fingerprint” i.e. the activation energy of thermal emission of charge carriers from the defect and its capture cross-section. DLTS measurements performed at various reverse bias voltages, $V_R$, and filling pulse voltages, $V_P$, applied to the junction, enable investigation of various parts of the junction. Moreover, from the dependence of the DLTS-signal amplitude on the filling pulse time, $t_P$, the kinetics for capture of charge carriers into the defect states can be determined, which enable to differentiate isolated point defects from extended defects, like dislocations and interface or surface defect states [8,22].

DLTS spectra for the heterojunctions grown on the GaAs substrate, measured under various reverse bias voltages and filling pulse voltages, are shown in Fig. 8. In the investigated heterojunctions the space charge region was predominantly located within the undoped CdTe absorber layer, between the heavy doped *n*-type CdTe and *p*-type ZnTe layers. In our DLTS spectra negative peaks correspond to majority carrier traps, whereas the positive ones are associated with minority carrier traps. In order to determine whether they are hole or electron traps, the type of conductivity of the undoped CdTe layer has to be known. High concentration of cadmium vacancies commonly present in CdTe, which give rise to shallow acceptor states in the energy gap, causes that intentionally undoped CdTe crystals are predominantly of *p*-type [15,23,24]. Moreover, two of the traps revealed from the spectra shown in Fig. 8, denoted as H1 and H2, correspond to the ones defined as the majority carrier traps, determined from our previous DLTS measurements performed for Schottky junctions prepared on N-doped *p*-type CdTe layers grown by MBE on GaAs substrate [25,26]. On the grounds of the above mentioned findings we have concluded that the undoped CdTe absorber layers in our *p*-ZnTe/*i*-CdTe/*n*-CdTe heterojunctions were of *p*-type. Thus, the negative peaks observed in the spectra presented in Fig. 8 could be assigned to hole traps, labeled H1, H2, H7 and H8. The positive peak, observed only under injection conditions, i.e. under forward-bias filling pulse (spectrum (c) in Fig. 8), has been assigned to an electron trap, called E3.

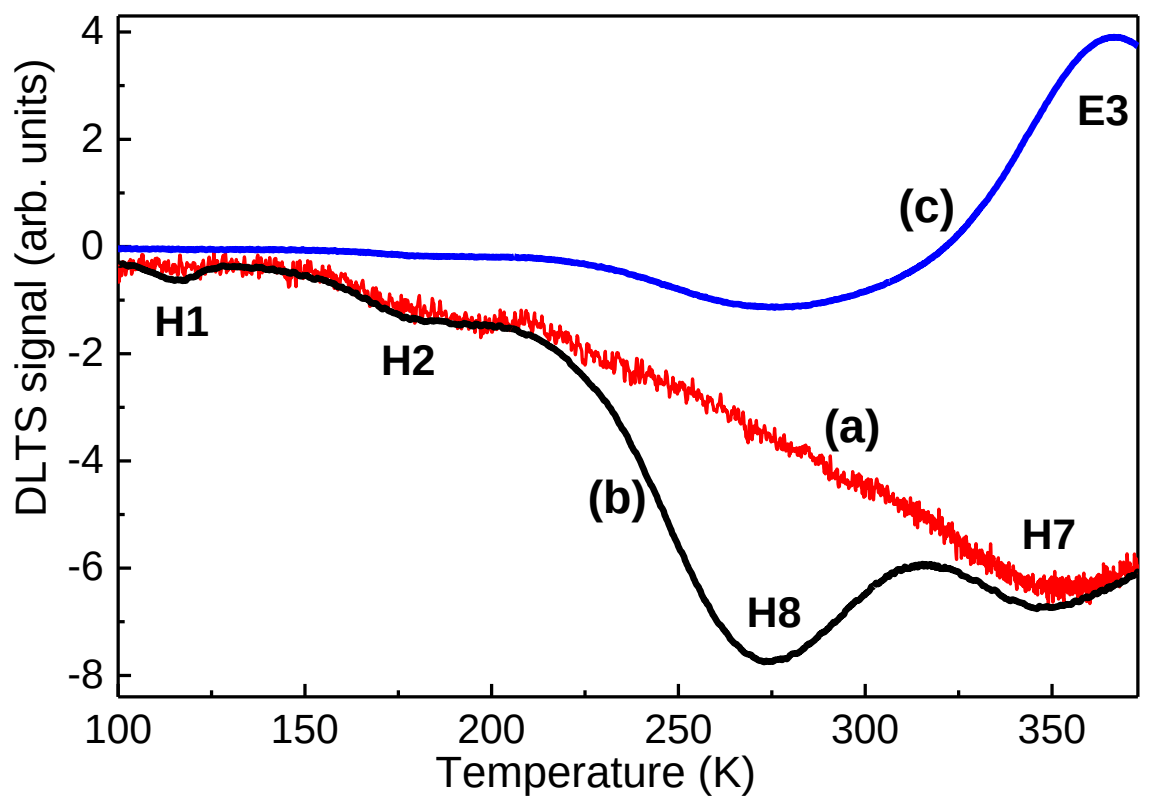

**Fig. 8.** DLTS spectra for the *p*-ZnTe/*i*-CdTe/*n*-CdTe heterojunction grown on the GaAs substrate measured at a rate window of 20.17 $s^{-1}$, filling pulse time $t_p$ = 1 ms and: $V_R = -3$ V and $V_P = -0.5$ V (a), $V_R = -3$ V and $V_P = 0.25$ V (b) and $V_R = -0.5$ V and $V_P = 0.5$ V (c).

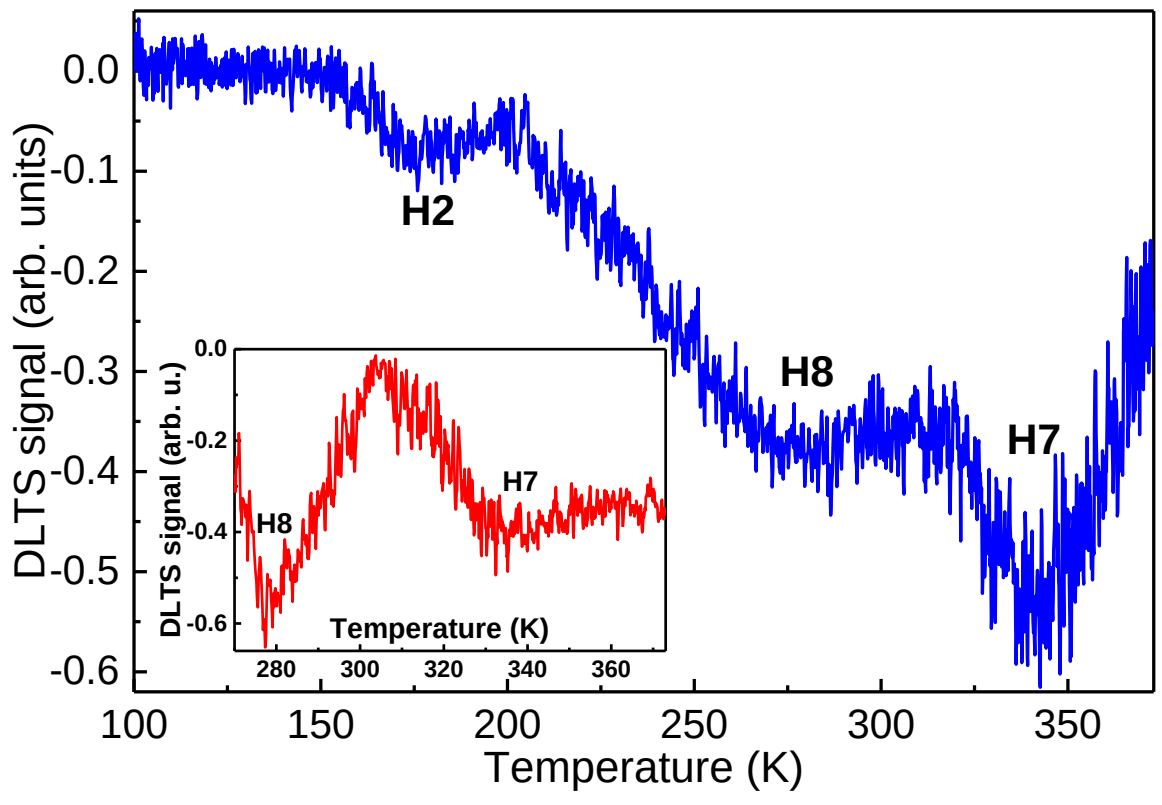

**Fig. 9.** DLTS spectra for the *p*-ZnTe/*i*-CdTe/*n*-CdTe heterojunction grown on the CdTe substrate measured at a rate window of 20.17 $s^{-1}$, filling pulse time $t_p$ = 1 ms and: $V_R = -3$ V and $V_P = 0.25$ V (main figure) and $V_R = -3$ V and $V_P = 1.5$ V (inset).

DLTS spectra representative for the *p*-ZnTe/*i*-CdTe/*n*-CdTe heterojunctions grown on the CdTe substrate are presented in Fig. 9. The H1 and E3 traps were not revealed in the homoepitaxially grown diodes. The positive DLTS peak of minority carrier trap E3, which, for the junction grown on GaAs, was clearly visible under the forward-bias filling pulse of $V_P = 0.5$ V (Fig. 8), was not observed for the junction grown on CdTe even at $V_P = 1.5$ V (inset in Fig. 9). The thermal emission rates of holes (electrons) from the traps, $e_p$ ($e_n$), as a function of reciprocal temperature (the Arrhenius plots), obtained for all the traps from temperature dependence of the DLTS spectra measured for various rate windows, are shown in Fig. 10. The emission activation energies of the traps, $E_T$, with respect to the valence, $E_V$, or conduction, $E_C$, band edge and their capture cross-sections, $\sigma$, evaluated from the slopes of the Arrhenius plots and their intersections with the ordinate, respectively, are listed in Table 2.

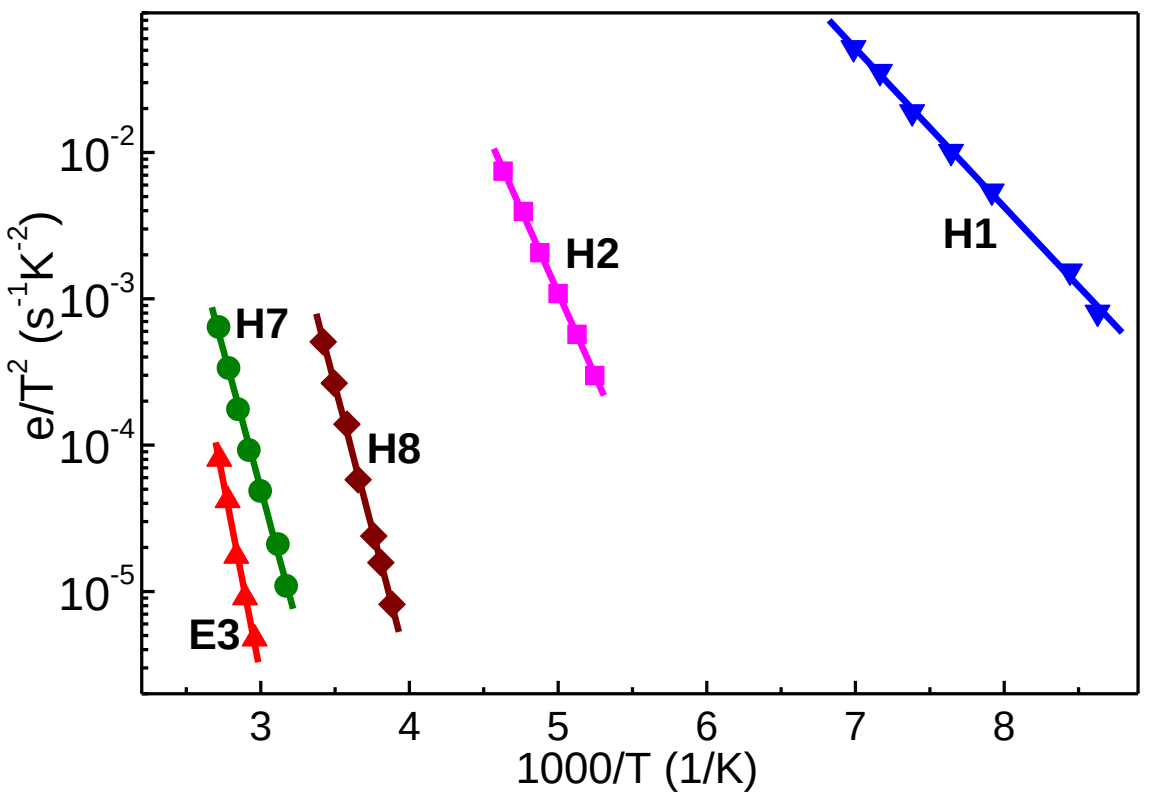

**Fig. 10.** Thermal emission rates as a function of reciprocal temperature (Arrhenius plots) for the traps revealed with DLTS in the *p*-ZnTe/*i*-CdTe/*n*-CdTe heterojunctions.

**Table 2.** Emission activation energies $E_T$, capture cross-sections $\sigma$ and proposed defects responsible for the traps revealed in the *p*-ZnTe/*i*-CdTe/*n*-CdTe heterojunctions.

| Trap | $E_T$ (eV) | $\sigma$ (cm$^2$) | Proposed defect |
|---|---|---|---|
| H1 | $E_V + 0.21$ | $1.35\times10^{-15}$ | dislocation states |
| H2 | $E_V + 0.45$ | $1.85\times10^{-13}$ | $V_{Cd}$ (2–/–) |
| H7 | $E_V + 0.75$ | $8.01\times10^{-15}$ | dislocation states |
| H8 | $E_V + 0.78$ | $8.10\times10^{-12}$ | dislocation core states |
| E3 | $E_C - 1.08$ | $3.75\times10^{-10}$ | dislocation core states |

The capture kinetics of charge carriers at the revealed deep-level traps are presented in Fig. 11. The dependence of the amplitude of DLTS peak assigned to the H2 trap on the filling pulse time shows a distinct saturation for longer filling times. It is characterized by the standard exponential capture kinetics [21], which has been fitted to the experimental points with the dashed line in Fig. 11. Such behavior is characteristic of isolated point defects or impurities. The same trap has been revealed as a majority carrier trap in our previous DLTS spectra obtained for MBE-grown *p*-type CdTe layers [25,26] and as a minority carrier trap in the spectra for MBE-grown iodine-doped *n*-type CdTe layers [27]. We have ascribed this hole trap to the (2–/–) level of Cd vacancy, $V_{Cd}$ [27], which gives rise to deep-level state at about $E_V + 0.45$ eV in the CdTe band gap [15,28], as distinct from the singly ionized charge state of Cd vacancy acting as a shallow acceptor state.

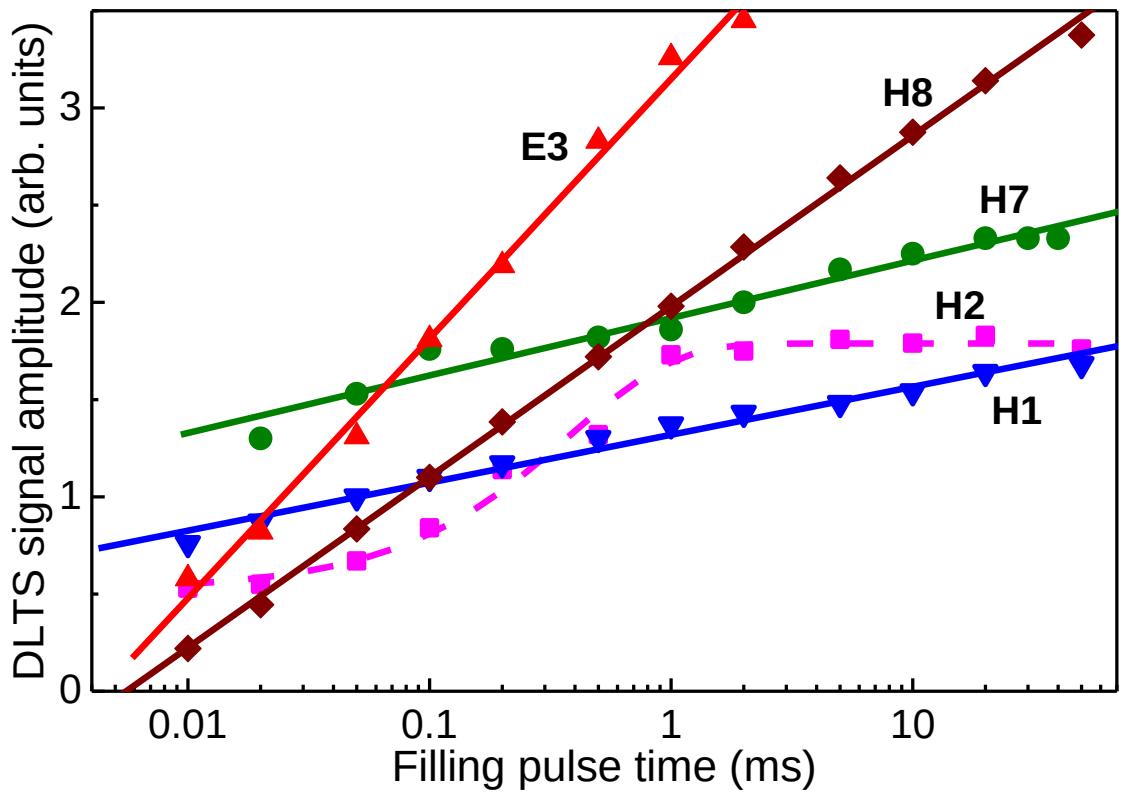

**Fig. 11.** Dependence of the DLTS-signal amplitude on the filling pulse time for all the revealed traps at the logarithmic time scale. The dashed and solid lines represent fittings of the exponential and logarithmic capture kinetics, respectively, to the experimental points.

On the other hand, the dependence of DLTS-peak amplitude on the filling pulse time, measured for all the other traps over almost four orders of magnitude of that time, was linear with the logarithm of that time, as shown in Fig. 11. Such dependence is a manifestation of logarithmic capture kinetics of charge carriers at the traps, which is characteristic of extended defects, such as dislocations, associated with the traps [29,30]. It results from the electrostatic interaction between a charge carrier just being captured and other charges already captured at the adjacent states of extended defect, which causes the formation of a Coulombic barrier around the charged extended defect [31]. The barrier height increases with the number of charge carriers captured at the defect, i.e. with the filling pulse time $t_P$ in DLTS experiment, and results in capture kinetics, which can be described by the following relation [30]:

$$n_T(t_P) = n\sigma\langle v_n\rangle N_T \tau \ln[(t_P + \tau)/\tau], \qquad (2)$$

where $n_T$ is the concentration of the traps filled with charge carriers, $n$ is the free carrier concentration, $\langle v_n\rangle$ is the charge-carrier thermal velocity, $N_T$ is the average trap concentration and $\tau$ is the time constant. According to this relation $n_T$ and, consequently, the DLTS peak amplitude increase linearly with logarithm of the filling pulse time (for $t_P > \tau$), similarly to those shown in Fig. 11 for the H1, H7, H8 and E3 traps.

Dislocations in semiconductors, displaying translational symmetry along dislocation lines, can give rise to the formation of one-dimensional energy bands in the energy gap rather than isolated electron states. Schröter et al. [32,33] proposed that electronic states associated with dislocations, characterized by logarithmic capture kinetics, can be distinguished as either so-called *localized* or *bandlike* states by investigation of the dependence of dislocation-related DLTS-peak shape on the filling-pulse time. The two types of states differ by the relation of the internal equilibration rate, $R_i$, at which the states within the defect energy band reach their internal electron equilibrium, to the carrier emission rate from the defect, $R_e$, and the capture rate, $R_c$.

For *localized* states: $R_i << R_e$, $R_c$, which results in random distribution of charge carriers trapped at the defect states during the filling pulse and stable position of the corresponding peak in the DLTS spectrum while increasing the filling pulse time. Its position on the temperature scale corresponds to the mean of the defect energy band. This is the case of both the H1 and H7 traps revealed in the present investigations. This type of electronic states has previously been ascribed to 60°-dislocations in plastically deformed silicon [32] and lattice-mismatch induced threading dislocations in the InGaAs/GaAs [34] and InGaAs/InP [35] heterostructures on the grounds of thorough analysis of the shape evolution of DLTS peaks associated with those dislocations while increasing the filling pulse time.

On the contrary, for *bandlike* states: $R_i >> R_e, R_c$. This relation results in occupying the lowest lying states in the defect energy band during the filling pulse. As a consequence, the occupation limit of the defect energy band increases, and the apparent emission activation energy of charge carriers from the defect decreases, while increasing the filling pulse time. This, in turn, causes that the corresponding DLTS peak broadens and its maximum shifts towards lower temperature with increasing filling pulse time. Such behavior is characteristic of the H8 hole trap (shown in Fig. 12) and of the E3 electron trap (Fig. 13). Previously, the *bandlike* states have been attributed to the electronic states associated with dislocation rings bounding nanoscale $NiSi_2$ precipitates in silicon [33] and to the core states of misfit dislocations at the InGaAs/GaAs [34] and InGaAs/InP [35] interfaces. The *bandlike* states are expected to be associated with the core states of "clean" dislocations, which are free of jogs, kinks and other imperfections. The *localized* states, on the other hand, are probably associated with dislocations containing defects in their cores or surrounded by point defect clouds [33].

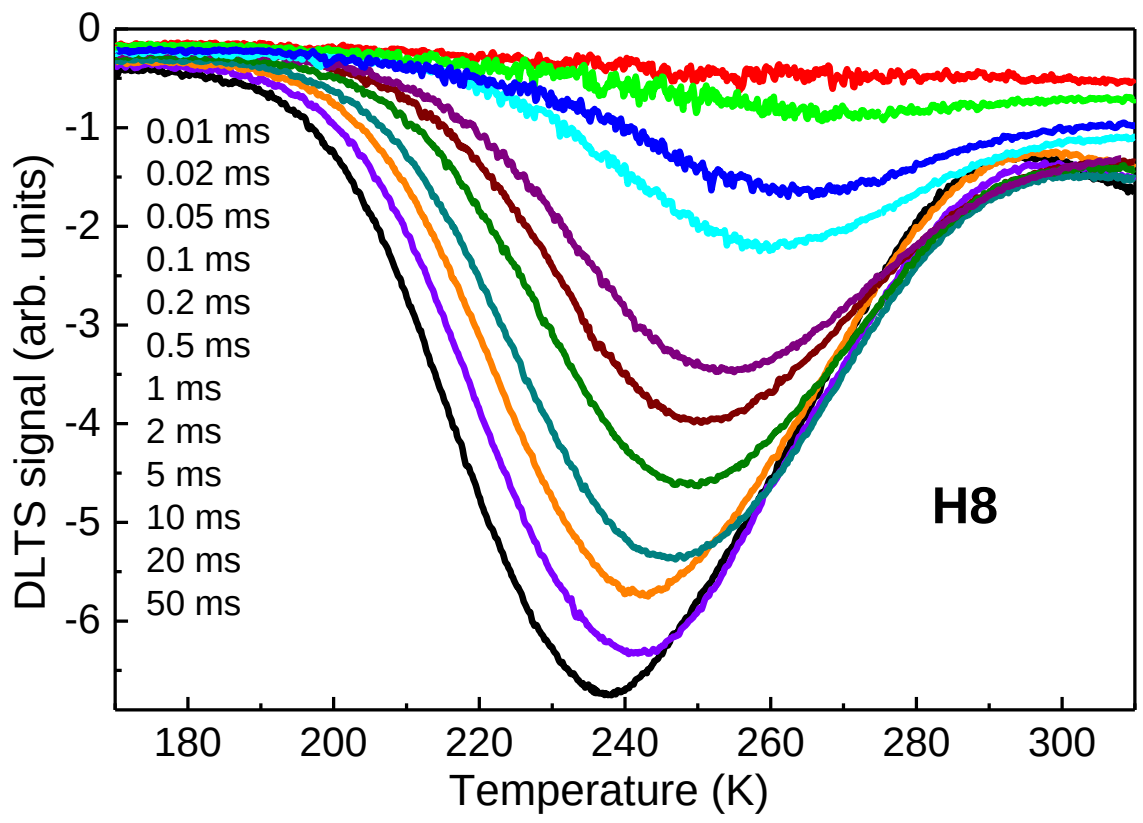

**Fig. 12.** Evolution of the DLTS-peak shape of the H8 hole trap while increasing the filling pulse time, whose values, in increasing order, are written in the figure.

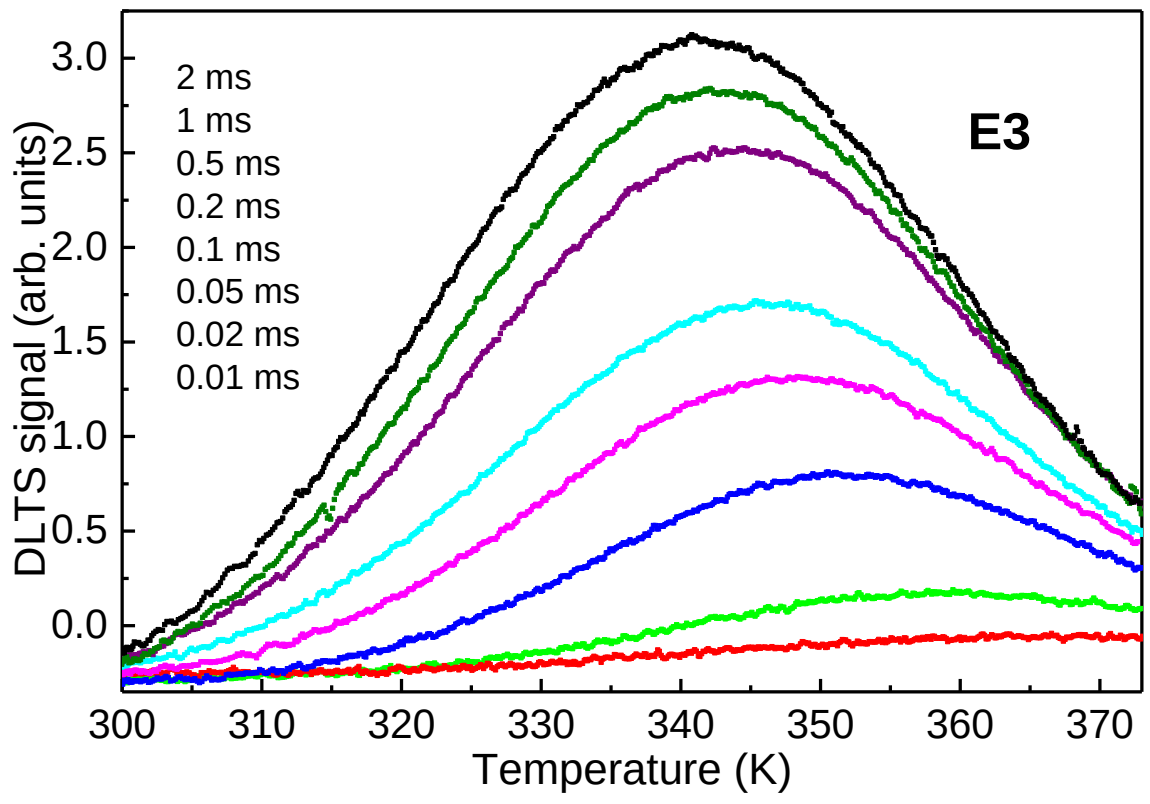

**Fig. 13.** Evolution of the DLTS-peak shape of the E3 electron trap while increasing the filling pulse time, whose values, in decreasing order, are written in the figure.

Three hole traps, H1, H7 and H8 and the electron trap E3, which exhibit logarithmic capture kinetics, are most likely related to dislocations revealed in the investigated heterojunctions by means of cross-sectional TEM imaging, shown in Figs. 3 and 4. Both the H1 and E3 traps were observed only in the heterojunctions grown on GaAs substrate. They were also revealed in DLTS measurements performed for Schottky junctions prepared on either *p*-type (H1 hole trap [25,26]) or *n*-type (E3 electron trap [27]) CdTe layers grown by MBE on GaAs substrate. They are probably associated with threading dislocations generated at the mismatched interface with the GaAs substrate and propagated through the CdTe layers during their growth. The H7 and H8 hole traps were observed in the heterojunctions grown on both GaAs and CdTe substrates. On the other hand, they were not revealed in the DLTS spectra of Schottky junctions prepared on MBE-grown CdTe layers [25,26]. Thus, they are probably related to dislocations created at the lattice mismatched ZnTe/CdTe interface. The hole trap, similar to the H7 one, displaying logarithmic capture kinetics and *localized* state behavior, was revealed in the DLTS spectra of Schottky junctions prepared on *p*-type CdTe polycrystals [36]. That trap, called H4 in Ref. [36], was ascribed by the authors to dislocations or clouds of point defects around dislocations. The proposed defects responsible for the deep-level traps revealed in our experiments are summarized in Table 2.

The H8 and E3 traps, ascribed to the core states of dislocations, are characterized by the largest values of the capture cross-section from all the traps revealed in the present investigations (see Table 2). This suggests that they may act as efficient recombination centers decreasing the carrier lifetime in the heterojunctions. These deep-level traps may primarily be responsible for rather poor energy conversion efficiency of below 5% determined for similar heterojunctions grown on GaAs substrate (without an antireflectance layer) from the measurements at 1-sun illumination [6,37]. Surprisingly, much higher values of the conversion efficiency, with the current highest value of 21% [38,39], have been obtained for solar cells based on polycrystalline CdTe. The dominant technological process allowing to achieve such high efficiency is the post-growth chloride activation treatment of polycrystalline CdTe layers [40,41]. This process consists in annealing the layers after exposure to $CdCl_2$ or $MgCl_2$ chlorides. Major et al. [42] have recently demonstrated that the chloride treatment, which increases the minority carrier lifetime and improves carrier transport in the solar cells, results in the passivation of CdTe grain boundaries by chlorine in-diffusion to the boundaries. In polycrystalline CdTe dislocations gliding through the crystalline grains become immobilized at the grain boundaries. We suggest that the electrical activity of these dislocations can be effectively diminished in the process of chloride treatment of polycrystalline CdTe layers as a result of chlorine incorporation at the grain boundaries.

## 4. Summary and conclusions

*p*-ZnTe/*i*-CdTe/*n*-CdTe heterojunctions, designed for photovoltaic applications, were grown by molecular-beam epitaxy on either GaAs or CdTe substrates and subjected to thorough examination by means of several complementary research techniques. High-resolution X-ray diffractometry results showed that the CdTe layers grown on GaAs under compressive lattice-mismatch strain were almost fully relaxed and displayed residual compressive strain of the order of $10^{-4}$ caused by different thermal expansion coefficients of CdTe and GaAs. The CdTe layers grown homoepitaxially on the CdTe substrate displayed much better structural quality and, unexpectedly, a small compressive strain, which results from imperfect structural quality of the CdTe substrate.

The cross-sectional TEM imaging of the CdTe/GaAs and ZnTe/CdTe interfaces revealed a high density of misfit dislocations at the interfaces, which act as a source of threading dislocations propagating through the whole thickness of both the CdTe and ZnTe layers. The geometry of threading dislocations suggests that most of them are of 60°-type. The high-resolution TEM images revealed also the presence of numerous stacking faults in the mismatch-strained CdTe and ZnTe layers.

Deep-level transient spectroscopy measurements revealed four deep-level hole traps, labeled H1, H2, H7 and H8, and one electron trap, labeled E3, in the heterojunctions grown on GaAs substrate and only three hole traps, H2, H7 and H8, in the heterojunctions grown on CdTe. The revealed traps are predominantly located within the undoped CdTe absorber layers in the heterojunctions. The H2 trap, displaying exponential capture kinetics of charge carriers, characteristic of isolated point defects, has been assigned to the double acceptor level of Cd vacancies produced during the epitaxial growth of CdTe layers. The other four traps, H1, H7, H8 and E3, which display the logarithmic capture kinetics, have been related to extended defects, presumably dislocations. The H1 and E3 traps, which were not observed in the DLTS spectra of heterojunctions grown on CdTe substrate, are probably associated with threading dislocations originated at the lattice-mismatched CdTe/GaAs interface. The H7 and H8 traps, on the other hand, observed in both the types of measured junctions, are likely related to extended defects created as a result of lattice mismatch between the measured CdTe absorber and the ZnTe layer. On the grounds of the dependence of DLTS-peak shape on the filling-pulse time we have ascribed the electronic states of the E3 and H8 traps to the core states of dislocations. Moreover, those two traps are characterized by the largest values of their capture cross-section, suggesting that they may act as efficient recombination centers decreasing the lifetime of photo-excited charge carriers in the heterojunctions and may primarily be responsible for a low energy conversion efficiency of photovoltaic devices based on similarly grown heterojunctions.

## Acknowledgment

This work was partly supported by the National Science Centre (Poland) by the grant No. UMO-2017/25/B/ST3/02966.

## References

[1] A. Louwen, W.G.J.H.M.A. van Sark, P.C.E. Faaij, I. Schropp, Re-assessment of net energy production and greenhouse gas emissions avoidance after 40 years of photovoltaics development, Nat. Commun. 7 (2016) 13728.

[2] Y. Wang, T. Muryobayashi, K. Nakada, Z. Li, A. Yamada, Correlation between carrier recombination and valence band offset effect of graded Cu(In,Ga)$Se_2$ solar cells, Sol. Energy Mater Sol. Cells 201 (2019) 110070.

[3] Y.A.R. Dasilva, R. Kozak, R. Erni, M.D. Rossell, Structural defects in cubic semiconductors characterized by aberration-corrected scanning transmission electron microscopy, Ultramicroscopy 176 (2017) 11-22.

[4] K. Wichrowska, J.Z. Domagala, T. Wosinski, S. Chusnutdinow, G. Karczewski, High-resolution X-ray diffraction studies on MBE-grown *p*-ZnTe/*n*-CdTe hetrojunctions for solar cell applications, Acta Phys. Polon. A 126 (2014) 1083-1086.
[5] K. Wichrowska, T. Wosinski, S. Kret, M. Rawski, O. Yastrubchak, S. Chusnutdinow, G. Karczewski, Extended defects in MBE-grown CdTe-based solar cells, Phys. Status Solidi C 12 (2015) 1115-1118.
[6] G. Karczewski, S. Chusnutdinow, K. Olender, T. Wosinski, T. Wojtowicz, Identification of recombination centers responsible for reduction of energy conversion efficiency in CdTe-based solar cells, Phys. Status Solidi C 11 (2014) 1296-1299.
[7] T. Wosinski, A. Makosa, Z. Witczak, Transformation of native defects in bulk GaAs under ultrasonic vibration, Semicond. Sci. Technol. 9 (1994) 2047-2052.
[8] K. Wichrowska, T. Wosinski, Z. Tkaczyk, V. Kolkovsky, G. Karczewski, Surface acceptor states in MBE-grown CdTe layers, J. Appl. Phys. 123 (2018) 161522.
[9] H. Tatsuoka, H. Kuwabara, H. Fujiyasu, Y. Nakanishi, Growth of CdTe on GaAs by hot-wall epitaxy and its stress relaxation, J. Appl. Phys. 65 (1989) 2073-2077.
[10] H. Tatsuoka, H. Kuwabara, Y. Nakanishi, H. Fujiyasu, Strain relaxation of CdTe(100) layers by hot-wall epitaxy on GaAs(100) substrates, J. Appl. Phys. 67 (1990) 6860-6864.
[11] D.G. Thomas, Excitons and band splitting produced by uniaxial stress in CdTe, J. Appl. Phys. 32 (1961) 2298-2304.
[12] H. Heinke, A. Waag, M.O. Möller, M.M. Regnet, G. Landwehr, Unusual strain in homoepitaxial CdTe(001) layers grown by molecular beam epitaxy, J. Cryst. Growth 135 (1994) 53-60.
[13] P.F. Fewster, N.L. Andrew, Absolute lattice-parameter measurement, J. Appl. Cryst. 28 (1995) 451-458.
[14] M. Leszczynski, J. Bak-Misiuk, J. Domagala, J. Muszalski, M. Kaniewska, J. Marczewski, Lattice dilation by free electrons in heavily doped GaAs:Si, Appl. Phys. Lett. 67 (1995) 539-541.
[15] A. Carvalho, A.K. Tagantsev, S. Öberg, P.R. Briddon, N. Setter, Cation-site intrinsic defects in Zn-doped CdTe, Phys. Rev. B 81 (2010) 075215.
[16] Y. Wang, P. Ruterana, S. Kret, S. El Kazzi, L. Desplanque, X. Wallart, The source of the threading dislocation in GaSb/GaAs hetero-structures and their propagation mechanism, Appl. Phys. Lett. 102 (2013) 052102.
[17] Y. Yan, M.M. Al-Jassim, K.M. Jones, S.-H. Wei, S.B. Zhang, Observation and first-principles calculation of buried wurtzite phases in zinc-blende CdTe thin films, Appl. Phys. Lett. 77 (2000) 1461-1463.
[18] J.D. Poplawsky, W. Guo, N. Paudel, A. Ng, K. More, D. Leonard, Y. Yan, Structural and compositional dependence of the $CdTe_xSe_{1-x}$ alloy layer photoactivity in CdTe-based solar cells, Nat. Commun. 7 (2016) 12537.
[19] A.R. Peaker, V.P. Markevich, J. Coutinho, Tutorial: Junction spectroscopy techniques and deep-level defects in semiconductors, J. Appl. Phys. 123 (2018) 161559.
[20] D.V. Lang, Deep-level transient spectroscopy – new method to characterize traps in semiconductors, J. Appl. Phys. 45 (1974) 3023-3032.
[21] D.V. Lang, Space-charge spectroscopy in semiconductors, in P. Bräunlich (Ed.), Topics in Applied Physics, Vol. 37, Springer, Berlin, 1979, pp. 93-133.

[22] T. Wosinski, O. Yastrubchak, A. Makosa, T. Figielski, Deep-level defects at lattice-mismatched interfaces in GaAs-based heterojunctions, J. Phys.: Condens. Matter 12 (2000) 10153-10160.

[23] A. Castaldini, A. Cavallini, B. Fraboni, Deep energy levels in CdTe and CdZnTe, J. Appl. Phys. 83 (1998) 2121-2126.

[24] S.J. Ikhmayies, R. N. Ahmad-Bitar, Characterization of vacuum evaporated CdTe thin films prepared at ambient temperature, Mat. Sci. Semicond. Proc. 16 (2013) 118-125.

[25] K. Olender, T. Wosinski, A. Makosa, P. Dluzewski, V. Kolkovsky, G. Karczewski, Native deep-level defects in MBE-grown *p*-type CdTe, Acta Phys. Polon. A 120 (2011) 946-949.

[26] K. Olender, T. Wosinski, A. Makosa, P. Dluzewski, V. Kolkovsky, Z. Tkaczyk, G. Karczewski, Extended deep-level defects in MBE-grown p-type CdTe layers, Phys. Status Solidi C 10 (2013) 113-116.

[27] K. Olender, T. Wosinski, A. Makosa, S. Kret, V. Kolkovsky, G. Karczewski, Capture kinetics at deep-level defects in MBE-grown CdTe layers, Semicond, Sci. Technol. 26 (2011) 045008.

[28] B.K. Meyer, D.M. Hofmann, Anion and cation vacancies in CdTe, Appl. Phys. A 61 (1995) 213-215.

[29] P. Omling, E.R. Weber, L. Montelius, H. Alexander, J. Michel, Electrical properties of dislocations and point defects in plastically deformed silicon, Phys. Rev. B 32 (1985) 6571-6581.

[30] T. Wosinski, Evidence for the electron traps at dislocations in GaAs crystals, J. Appl. Phys. 65 (1989) 1566-1570.

[31] T. Figielski, Recombination at dislocations, Solid State Electron. 21 (1978) 1403-1412.

[32] W. Schröter, J. Kronewitz, U. Gnauert, M. Seibt, Band-like and localized states at extended defects in silicon, Phys. Rev. B 52 (1995) 13726-13729.

[33] W. Schröter, H. Hedemann, V. Kveder, F. Riedel, Measurements of energy spectra of extended defects, J. Phys.: Condens. Matter 14 (2002) 13047-13059.

[34] O. Yastrubchak, T. Wosinski, A. Makosa, T. Figielski, A.L. Tóth, Capture kinetics at deep-level defects in lattice-mismatched GaAs-based heterostructures, Physica B 308-310 (2001) 757-760.

[35] P.-C. Hsu, E. Simoen, C. Merckling, G. Eneman, Y. Mols, A.R. Alian, R. Langer, N. Collaert, M. Heyns, Bandlike and localized states of extended defects in n-type $In_{0.53}Ga_{0.47}As$, J. Appl. Phys. 124 (2018) 165707.

[36] A.E. Bobrova, Yu.V. Klevkov, S.A. Medvedev, A.F. Plotnikov, A DLTS study of deep levels in the band gap of textured stoichiometric *p*-CdTe polycrystals, Semiconductors 36 (2002) 1426-1431.

[37] E. Zielony, K. Olender, E. Placzek-Popko, T. Wosinski, A. Racino, Z. Gumienny, G. Karczewski, S. Chusnutdinow, Electrical and photovoltaic properties of CdTe/ZnTe *n-i-p* junctions grown by molecular beam epitaxy, J. Appl. Phys. 115 (2014) 244501.

[38] M.A. Green, K. Emery, Y. Hishikawa, W. Warta, E.D. Dunlop, D.H. Levi, A.W.Y. Ho-Baillie, Solar cell efficiency tables (version 49), Prog. Photovolt. 25 (2017) 3-13.

[39] O.S. Ogedengbe, C.H. Swartz, P.A.R.D. Jayathilaka, J.E. Petersen, S. Sohal, E.G. Leblanc, M. Edirisooriya, K.N. Zaunbrecher, A. Wang, T.M. Barnes, T.H. Myers, Iodine doping of CdTe and CdMgTe for photovoltaic applications, J. Electron. Mater. 46 (2017) 5424-5429.

[40] C. Li, J. Poplawsky, Y. Wu, A.R. Lupini, A. Mouti, D.N. Leonard, N. Paudel, K. Jones, W. Yin, M. Al-Jassim, Y. Yan, S.J. Pennycook, From atomic structure to photovoltaic properties in CdTe solar cells, Ultramicroscopy 134 (2013) 113-125.
[41] J.D. Major, R.E. Treharne, L.J. Phillips, K. Durose, A low-cost non-toxic post-growth activation sStep for CdTe solar cells, Nature 511 (2014) 334-337.
[42] J.D. Major, M. Al Turkestani, L. Bowen, M. Brossard, C. Li, P. Lagoudakis, S.J. Pennycook, L.J. Phillips, R.E. Treharne, K. Durose, In-depth analysis of chlorine treatments for thin-film CdTe solar cells, Nat. Commun. 7 (2016) 13231.